*Article*

# E-polis: Gamifying Sociological Surveys through Serious Games - A Data Analysis Approach Applied to Multiple-Choice Question Responses Datasets

**Alexandros Gazis** [1,*] **and Eleftheria Katsiri** [1]

[1] Department of Electrical and Computer Engineering, School of Engineering, Democritus University of Thrace, 67100 Xanthi, Greece;
* Correspondence: agazis@ee.duth.gr

**Abstract:** E-polis is a serious digital game designed to gamify sociological surveys studying young people's political opinions. In this platform game, players navigate a digital world, encountering quests posing sociological questions. Players' answers shape the city-game world, altering building structures based on their choices. E-polis is a serious game, not a government simulation, aiming to understand players' behaviors and opinions thus we do not train the players but rather understand them and help them visualize their choices in shaping a city's future. Also, it is noticed that no correct or incorrect answers apply. Moreover, our game utilizes a novel middleware architecture for development, diverging from typical asset-prefab-scene and script segregation. This article presents the data layer of our game's middleware, specifically focusing on data analysis based on respondents' gameplay answers. E-polis represents an innovative approach to gamifying sociological research, providing a unique platform for gathering and analyzing data on political opinions among youth and contributing to the broader field of serious games.

## 1. Introduction

E-polis represents an innovative approach to conducting sociological surveys, taking the form of a digital game designed to engage participants in navigating a virtual city, [1]. The game's structure involves completing a predetermined set of quests, which serve as pathways to advance to higher levels or conclude the gaming experience. These quests are meticulously crafted by political scientists and public administration officers affiliated with the National and Kapodistrian University of Athens (NKUA), [1]. The scenarios and questions presented in the game cover a broad spectrum of political profiles, spanning from the right to the left wing. In Figure 1 and Figure 2 we illustrate the actual gameplay of the first level of our game:

**Figure 1.** The first level of our game is where in this view we can see the graphics.

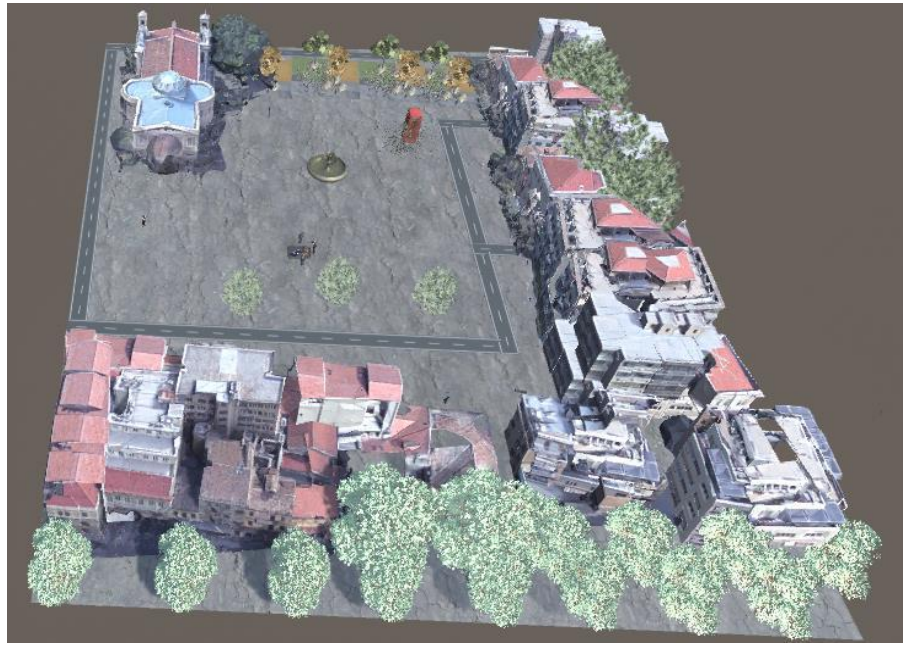

**Figure 2.** The first level of our game is from a different view.

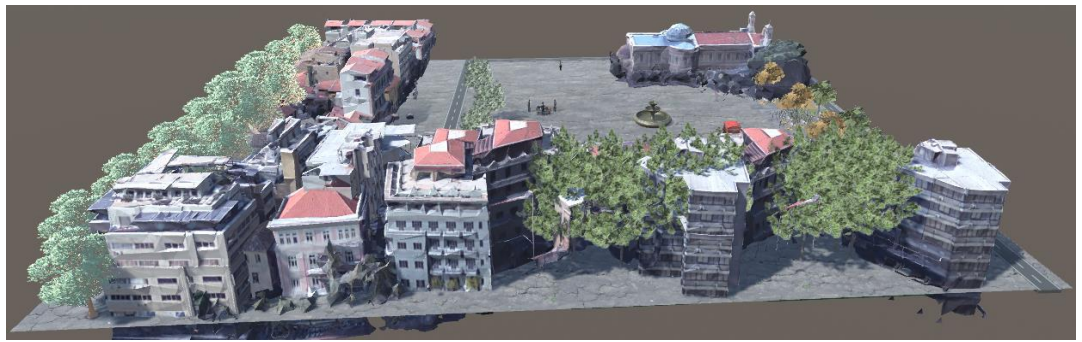

It is essential to note that E-polis falls under the category of serious games, where the primary objective is to leverage the entertainment aspects inherent in digital games to encourage active participation. As an educational game, E-polis currently lacks a scoring system, and there is no specific time frame imposed for completing each level or scene. The emphasis remains on utilizing the engaging nature of digital games to stimulate participants' involvement, [2], in the sociological survey presented within the gaming environment, [3]. A key emphasis is placed on delving into the transformative potential inherent in the imaginative and utopian thoughts of young people. Notably, the project introduces a groundbreaking element by creating a digital environment in the form of a video game. This innovative approach allows for the examination of players' reactions and preferences under simulated conditions. The utilization of digital games as methodological tools for social research introduces fresh, transdisciplinary approaches to the field. The analysis of digital worlds, where participants freely engage with social and political issues unbound by physical reality, yields valuable investigative data. This signifies a departure from traditional research methodologies, providing a deeper understanding of how individuals respond to and interact with simulated scenarios within the context of the E-polis project, [4].

Building upon our prior study, [5], we propose the adoption of a novel middleware for the development of a digital game that emphasizes a targeted didactic or pedagogic approach. In the context of our discussion, middleware is defined as an abstract software entity designed to integrate various components and functionalities within software systems. Specifically, our middleware recommendation entails the following separation of concerns, [5]:

- Functional Tasks Integration (Platform Layer):

  The Platform layer primarily handles the actual gameplay, distinct from the utilization of available hardware and the creation of executable files for various platforms and operating systems. It ensures the seamless integration of functional tasks within the game.

- Process Coordination (Engine Layer):

  The Engine layer, encompassing suggested scene transition mechanisms, is where we pinpoint the coordination of different services and common engine procedures. It incorporates scene management, navigation, rendering, physics, and other modules and components vital to the game's engine functionality.

- Endpoint Provision (Game Layer):

  The Game layer, central to the focus of this article, specifically addresses the data analysis aspect. This layer is responsible for providing necessary services unique to each game, such as character controls and actual gameplay mechanics related to navigation, quests, and overall gameplay dynamics.

- Separation of Concerns (Application Layer):

  The Application layer, this aspect involves defining the game object hierarchy and its tight connection with scripts. It serves as a means for user interface input handling and network communication endpoints between the player and the game engine. The Application layer ensures a clear separation of concerns, contributing to a well-structured and maintainable system.

Furthermore, aside from the game's entertainment and educational aspects, a noteworthy technical innovation lies in our approach to transitioning between different scenes or levels using the lazy loading technique, [6,7]. This method aims to optimize the necessary computing resources for playing our game by dynamically managing the loading of graphics based on the player's navigation and field of view, [1].

To illustrate, when a player is oriented towards the north, we selectively load only the corresponding part of the game, employing threads to efficiently offload graphics from memory for the current level. The primary objective is to minimize resource consumption and enhance performance during gameplay, [1]. An essential design decision involves limiting the game to a singular light source – the sun. This choice, excluding additional rays, is intentional due to the resource-intensive nature of rendering and shading processes during both gameplay and the final scene rendering. This streamlined approach ensures a focused and optimized use of computational resources throughout the gaming experience and when constructing the ultimate scene of the game.

Regarding the theoretical innovation embedded in our game, beyond the middleware layer, we have introduced a "do not repeat yourself" (DRY) approach, [5], in both front-end and back-end operations during game development. Additionally, we have elucidated how to leverage a pre-existing game engine, Unity, to build upon and expand functionality, [8-10].

In this article, our focus extends beyond presenting the actual game or each layer's functionality. Instead, it centres on the data layer, specifically emphasizing the analysis – the data transformation part – of the serious game named E-polis. The subsequent sections will detail the provided questions and the rationale behind constructing a database with various types of questions based on sociological profiles. We will then delve into the Python programming language scripts and algorithms employed to generate dummy data, as actual data is restricted due to GDPR, and the game is still in its developmental phase.

Notably, the axis of our analysis is presented in Greek, aligning with the final deliverable intended for submission in the Greek language to the Hellenic Foundation of Research and Innovation (HFRI)[1].

In conclusion, we summarize our findings, emphasizing that this initial analysis serves as a validation tool for the ground truth, analytically checking and validating specific quality traits or behaviours exhibited by participants. The outcomes of this research will contribute valuable insights to the ongoing development of the E-polis serious game.

## 2. Related Works

### 2.1. Brief Literature Review

A serious game is a game whose primary design and focus are to educate, train, or raise awareness, rather than entertain players. This means that its design principles centre around using interactive elements to achieve objectives beyond entertainment, typically related to training individuals or helping them learn. As such, these games incorporate elements of gamification but are usually not highly competitive and do not force correct or incorrect answers on players, though they can include a points or reward system, [11-13].

The purpose of an educational -serious- game is to provide a simple, engaging experience that does not require extensive computer resources or complex graphics and animations, [14,15]. Educational games, such as the one discussed here, typically prioritize gameplay and learning outcomes over high-end visuals or sophisticated rendering mechanisms, [16-19]. While there are numerous game engines available, each with its strengths and weaknesses, many serious games in recent literature utilize pre-existing solutions, [20-23], simple Web-based frameworks, [24,25], or cloud-hosted solutions, [26-28], making them lightweight and easy to deploy.

The key consideration in such educational games is the gameplay itself, [29,30]. Specifically, whether the game demands live, quick, and interactive player engagement or a more relaxed approach, where players mainly read text dialogues and select answers, [31,32]. The recent trend has been to use JavaScript frameworks or convert the game into a WebGL version, as it is important to quickly develop a prototype for immediate play and then iterate on its features based on player feedback, [33-37].

The digital game industry follows a Rapid Application Development (RAD) approach, which emphasizes fast feedback from players rather than focusing on a fixed release plan for features. This iterative approach allows for the release of multiple updates, each focused on enhancing quality and adding new features throughout the development process. Prototypes are created based on user design requirements and refined continuously as player interaction shapes the game's development. This method ensures the creation of agile, flexible, and scalable applications, [38-41].

The E-polis digital serious game discussed in this article stands out due to its unique research focus and comprehensive multimethodological approach, [1]. Rather than simply investigating young people's attitudes toward socio-political matters, [42-44], our digital game approach takes it a step further by actively encouraging participants to reconstruct their interests in political involvement, engage in debates around existing or proposed institutions, and imagine alternative forms of collective organization, [45,46]. This approach contributes to a redefinition of the concept of democracy, emphasizing the transformative potential inherent in young people's imaginative and utopian thoughts, [47-49]. This kind of project introduces a groundbreaking element by creating a virtual environment in the form of a video game, allowing for the examination of players' reactions and preferences under simulated conditions. The use of digital games as methodological tools in social research introduces fresh, transdisciplinary approaches to the field. By analyzing digital worlds where participants freely engage with social and political issues unbound by physical reality, the project generates valuable data and offers a deeper understanding of how individuals respond to and interact with simulated scenarios, [50-53].

---

[1] Hellenic Foundation of Research and Innovation (HFRI). Homepage. HFRI Website **2024**. Available online: https://www.elidek.gr/en/homepage (accessed on 19 May 2025).

As such it is important to promote the adoption of a novel middleware for the development of a digital game that focuses on a targeted didactic or pedagogical approach, [54,55]. This means that the term middleware, in this context, is defined as an abstract software entity designed to integrate various components and functionalities within software systems. Unlike monolithic applications, the proposed middleware does not operate as a singular entity but rather amalgamates distinct tasks and processes. It offers essential endpoints and a separation of concerns for each layer, ensuring seamless communication between them. This approach results in a modular and well-organized system, [56].

*2.2. Questions Presented During Gameplay*

The questions presented during each gameplay session are rooted in specific dilemmas. In this context, as the player navigates the city, they encounter real-life scenarios, such as an altercation between a homeless man and law enforcement officers. When the player approaches these incidents, a pop-up message is triggered. A dialogue textbox appears, presenting a question and a set of available responses. This mechanism is initiated using Unity's prefab functionality, [57], and is activated when the player enters the collision range of the prefab object, [58,59].

Each response corresponds to a particular political behaviour. The .csv file provided by political scientists specifies the location where each quest is designed to appear (e.g., city square, flea market) and the potential outcomes. It's important to note that questions are not presented with correct or incorrect options; instead, players are confronted with situations that require reflection and decision-making. To complete a game level, players must answer all the questions provided, [1,5]. If the game is stopped, exited, or interrupted in any way, meaning that the full range of answers for each question is not available, those samples are not considered for analysis.

There are two types of questions, and based on the players' answers, the graphics of our game change accordingly. The first group of dilemmas can be categorized as follows (Political Philosophy Dilemmas):

- <u>Democratic Radicalism</u>: Seeks societal transformation through democratic means.
- <u>Critical Liberalism</u>: Emphasizes social justice, and critiques traditional liberal thought.
- <u>Depoliticization</u>: Removes issues from the public sphere, exclusive to experts or elites.
- <u>Conservatism</u>: Emphasizes tradition, order, and stability.
- <u>Authoritarianism</u>: Strict control suppresses dissent in government.
- <u>Nihilism</u>: Rejects accepted aspects of human existence (knowledge, morality, etc.) represented by the Greek slang word "kava."

The second group of dilemmas can be categorized into the following (International Relations and Political Theory Dilemmas):

- <u>Realism</u>: Emphasizes power, national interest, and balance of power in international politics.
- <u>Technocracy</u>: Advocates rule by experts, particularly scientists and engineers.

- Cultural Reductionism: Believes cultural differences can be explained by a single factor like race, ethnicity, or religion.
- Humanism: Emphasizes human reason, freedom, and dignity in philosophy and ethics.
- Meritocracy: Rewards based on ability and effort, not social class or background.
- Communalism: Political and economic system based on cooperation, mutual aid, and shared resource ownership.

*2.3. Research Questions*

Both the actual game developed and the game middleware architecture expands upon existing design principles to present a top-down approach for creating serious games (SGs) intended for data collection on themes such as social justice, economic development, and the promotion of civic engagement and critical thinking among youth. Positioned as a modern research tool, this SG enables players to explore socio-political issues within a democratic context by reflecting on their in-game decisions. It not only collects user responses and behavioural data but also captures individual perspectives, decisions, and reactions concerning political involvement and societal operations. The technical innovation introduced through this game includes a middleware architecture that emphasizes modular software entities, behaviour, and interactions, structured through a clear separation between platform, engine, game, and application layers. This design supports both front-end and back-end development processes. Additionally, we propose a novel mechanism for scene transitions inspired by the lazy loading method to improve gameplay responsiveness.

From this foundation, our manuscript explores the following research questions:

1. **How can modular middleware architectures improve the performance of serious games?**
   - Existing middleware often introduces computational overhead, affecting rendering and simulations. We address this by decoupling front-end rendering from game logic and by implementing a lightweight scene transition mechanism to enhance performance with minimal memory usage.
   - Our research considers how cloud gaming and AI-powered engines may further improve middleware efficiency, proposing a contribution in the form of a modular, lightweight design for better integration.

2. **In what ways can serious games be optimized to collect behavioural data in real-time without compromising system performance or user privacy?**
   - Traditional middleware is not tailored for real-time analytics. Our approach integrates real-time event tracking and cloud-based storage using Firebase, ensuring both scalability and minimal performance impact.
   - The contribution lies in enabling secure and ethical behavioural data collection while preserving player anonymity.

3. **How can game middleware be designed for seamless cross-platform compatibility?**
   - Many middleware systems fail to support efficient deployment across platforms like PC, mobile, WebGL, and VR. We address this by separating front-end and back-end processes, allowing easy adaptation.

- This work contributes a platform-agnostic middleware design, suitable for integration with multiple game engines and deployment environments.

## 3. Methodology

*3.1. Game Resouces Used and Challenges*

3.1.1. DB tables

Initially, after successfully opening and loading the .csv question structure into our algorithmic system input, our focus shifted towards implementing a robust and agile solution—specifically, exporting our data into a single database instance. Our initial approach involved creating a local network within the development environment of the game and hosting a database on one of the local computers in our lab. Several free and open-source software solutions, such as Xampp, were considered for rapid prototyping and development. Xampp offered an inclusive framework, providing both a server for hosting our database (Apache) and a graphical user interface (PhpMyAdmin) for database interaction, [60]. However, integrating a connection with the Unity game platform proved challenging, as Xampp's solution was primarily designed for web applications.

Given Unity's usage of C# and the necessity to analyze and mine data using Python, and SQLite, "a C-language library that implements a small, fast, self-contained, high-reliability, full-featured, SQL database engine", [61] (version 3.39.2[2]). After extracting the file, we added it to the Unity assets layer, along with the necessary .dll files for the SQLite implementation on our local machine to efficiently manage and analyze player data using external modules such as [62,63].

Based on the information provided, during players' gameplay, we instantiate a script that incorporates two functions for a database:

- <u>Create Table Function</u>: This function is designed to create a table with a specific structure to store players' answers if it does not already exist. It ensures that the database is appropriately configured to store the required information.
- <u>Insert Into Function</u>: The second function involves executing insert commands to add new rows to one of the two tables based on the player's quests. This function is responsible for populating the database with relevant data corresponding to the player's actions and choices during gameplay.

Each function operates as a standalone solution by opening and closing a connection to the database when invoked by the rest of our game. This modular approach ensures efficient and isolated functionality. Whenever a significant action occurs in the game, we invoke the script, triggering the execution of these functions, which in turn insert the necessary values into the respective database tables. This systematic integration of the script with the game's mechanics enables seamless data management and storage throughout the gameplay experience.

3.1.2. Software Specifications (Serious Game Engine Explained)

Firstly, when we started implementing our game, we needed to find a way for all players to interact within the same scene exclusively through the objects in the environment. This meant the multiplayer solution we chose had to be lightweight and avoid the need for complex systems. Since players in our game do not communicate via chat or microphone and only influence each other's decisions through changes in the buildings caused by responses to dilemmas, we prioritized simplicity and efficiency.

---

[2] SQLite. SQLite Documentation 2025. Available online: https://www.sqlite.org (accessed on 20 May 2025).

Given that our game's graphical requirements are minimal, i.e., low-polygon city models and low computational demand, we needed a quick and reliable network solution for rapid prototyping. Consequently, we researched existing Unity multiplayer frameworks and found an abundance of options, including:

We excluded simpler micro-frameworks like RakNet and WebSocketSharp, as well as remote database solutions like Firebase, which rely on REST APIs for multiplayer interactions. Our focus was on solutions specifically designed for Unity. While Firebase is popular in the C# community, its primary use in Unity is for login systems rather than full-game logic. Similarly, RakNet is better suited for resource-intensive games requiring high scalability, which doesn't align with our needs. WebSocketSharp, commonly used for low-latency applications, is often implemented in more complex, high-performance games.

Based on our requirements, we initially focused on three lightweight options that offered easy setup and scalability for our demo prototype: NetCode for Game Objects, Photon, and Mirror. After evaluating these, we found Photon and Mirror to be the most straightforward to integrate and use. Both are free and highly reliable for indie and AAA game development. In our opinion, Photon is better suited for small-scale games due to its built-in chat rooms and lobby system, while Mirror is more appropriate for large-scale multiplayer games with extensive tools for scalability.

Ultimately, we selected Photon because of its comprehensive documentation, active community, and support for future scalability. Photon provided the necessary tools to handle more users, increase complexity, and extend game features over time.

Our multiplayer implementation involved servers, clients, dedicated servers, and host servers. Specifically, a server acts as an instance of the game that all players connect to for shared gameplay. It manages various aspects, such as storing responses to dilemmas and transmitting data back to clients. Clients, on the other hand, are instances of the game running on individual devices, connecting to the server over a local or online network.

The server can either be a dedicated server or a host server. A dedicated server runs solely to manage connections and data, while a host server doubles as a player and a server. In our game, the server computer also allows users to play the game, making it a host server responsible for storing information and initializing the lobby.

For multiplayer games in Unity, the server must spawn game objects and synchronize changes across all players' instances. When the server spawns game objects, it ensures all connected clients replicate these objects. The spawning system manages the lifecycle of the objects and synchronizes their states. Photon handles this by associating each player's connection with a unique game object, ensuring only the respective player can directly modify their object. All changes are synchronized across the network, so the game world remains consistent for all players.

In Photon, the concept of "authority" determines control over game objects. By default, the server holds authority over all game objects, except for player-specific objects, which are managed with "local authority."

In our configuration, we used a host server and implemented a lobby menu at game startup. Players can create or join a room, enabling multiplayer functionality. Once all clients press the ready button, the host server starts the game. Each room supports up to six players, including the host, to maintain optimal performance and network responsiveness. Testing revealed that exceeding six players caused throttling in CPU performance and network response times.

To evaluate multiplayer features, we created multiple builds of the project to test functionality across devices. Using Photon, all builds must share the same ID to initialize properly. Thus, we focused on creating executable (.exe) files for x86 Windows systems and conducted tests on Windows 10 and 11 devices, as well as touchscreen tablets running Windows. These tests confirmed the game operated smoothly.

This workflow significantly expedited development, as we focused on optimizing the game for Windows. However, we plan to expand the project to Android and iOS architectures, enabling the game to run on Unix systems. This will allow us to package the game as an APK or VR application in the future with minimal adjustments.

As such, to play the game, the only resource needed is a computer device with an Internet connection to log into our multiplayer game.

3.1.3. Technical Challenges and Solutions

Middleware plays a foundational role in serious game development by bridging core engine functions, user interaction, and data management. However, existing middleware solutions often rely on monolithic or tightly coupled designs that are not suggested for real-time performance. These legacy architectures introduce computational overhead during rendering and physics simulations, limiting responsiveness. To address this, our middleware uses a more modular structure that separates front-end rendering from game logic execution, enabling more efficient resource allocation and performance optimization. We also implement a lightweight scene transition mechanism that supports dynamic content updates without consuming excessive memory—an important consideration for resource-limited environments such as mobile or browser-based games.

Beyond performance, a major gap in conventional middleware lies in the handling of behavioural data. While many serious games collect player interaction logs, traditional middleware does not support real-time analytics or scalable data storage. Our approach integrates real-time event tracking and cloud-based storage via Firebase, enabling the secure, anonymous collection and analysis of player decisions.

Scalability remains another critical limitation of existing middleware frameworks, particularly in multiplayer or data-intensive environments. Traditional client-server architectures restrict real-time responsiveness, as clients must send explicit requests before receiving server responses. Our middleware replaces this model with an event-driven architecture inspired by IoT systems, treating each player interaction as a continuous data stream. This shift enables decentralized, asynchronous communication, reducing latency and supporting seamless synchronization across clients. It also allows for dynamic scene updates without constant back-and-forth communication with a central server.

In terms of adaptive gameplay, most middleware solutions collect behavioural data passively but do not act on it in real time. Our middleware overcomes this by treating the game as an intelligent data-generating system. We incorporate adaptive architectural principles, using behavioural data to modify game scenes dynamically. This is achieved through AI-driven procedural content generation, enabling personalized experiences and greater player immersion.

Finally, conventional engines such as Unity lack built-in support for scalable database integration. Our middleware addresses this by incorporating principles from large-scale IoT architecture, enabling seamless integration with external databases for efficient player data storage and retrieval. By combining modular design, real-time data streaming, adaptive scene control, and cloud-native scalability, our middleware provides a comprehensive solution for the next generation of serious games, aligning with the demands of both developers and data-driven gameplay environments.

*3.2. Algorithms and Rationale of E-polis digital game*

Each function of the digital game operates independently, opening and closing a database connection whenever invoked by the game. This modular approach ensures efficiency and isolation. Whenever a significant in-game action occurs, the script is triggered, executing these functions to insert the necessary values into the appropriate database tables. This seamless integration between the script and the game's mechanics enables efficient data management and storage throughout gameplay.

3.2.1. Scene Transition Mechanism

The scene transition mechanism in *E-Polis* is responsible for creating a dynamic and interactive game environment where player choices influence and shape the game world's structure. As such, it generates a responsive and evolving environment. Unlike conventional static games, it modifies the game world in real time based on user input, accommodating up to six players per room in multiplayer mode.

3.2.2. Initial Algorithm for Scene Transition Using Prefab Collisions

In the early stages of our game's development, scene transitions were managed using a prefab-based collision detection system. Specifically, this technique relied on placing invisible player-trigger zones throughout the game world. When the player enters one of these zones, a question or dilemma will be triggered. This mechanism served as the core interaction for advancing the player through the game, as it linked physical navigation with actual decision-making. The system was designed not only to ensure that players encountered specific scenarios in a randomized sequence—requiring them to complete or respond to as many as possible within a given timeframe—but also to integrate the player's navigation with the evolving state of the game world based on their decisions.

In this section, we present a top-down overview of how this system functioned, its purpose, and how it laid the foundation for more advanced rendering and logic using the Unity game engine.

- Core principle:
  - Detects when a player enters a specific area.
  - Presents a dilemma (question) that the player must respond to.
  - Records the player's choice and adjusts game world variables.

- Purpose of implementation:
  - Used to trigger dilemmas based on player movement to a prefix game world space (road).
  - Ensures game progression only occurs when questions are answered.
  - Enables real-time interaction with the game environment variables.

- How the system works:
  - When a player enters a prefab area, the algorithm activates a pop-up containing a question (dilemma).
  - Player selection updates the game world variables and stores their response for analysis.

- o The dilemma does not disappear when answered, the player is moved outside of the trigger area to see the results of his/her choice and can re-enter and re-answer.
- Pseudocode Presentation:

```
def on_player_enter_area(player, prefab_area):
    if player.position in prefab_area:
        dilemma = get_dilemma(prefab_area)
        display_dilemma(dilemma)
        response = player.respond_to_dilemma()
        store_response(player.id, dilemma.id, response)
        update_scene_state(dilemma, response)
```

This was later expanded to use graphic rendering algorithms instead of prefabs.

Similarly, to improve scalability and visual immersion, the initial prefab-based collision system was later expanded to incorporate dynamic rendering algorithms. This transition enabled more flexible and visually coherent scene changes by leveraging Unity's real-time rendering capabilities. The algorithm for *the Advanced Scene Transition with Dynamic Rendering* was designed as follows:

- Core principle:
    - o Implements real-time changes to game graphics based on player decisions.
    - o Unlike the previous approach, pre-rendered objects are modified dynamically.
- Purpose of implementation:
    - o Enhances the visual part of our game by allowing players to construct the city's layout based on their responses.
    - o Avoids the limitations of prefabs by using modular rendering techniques.
- How the system works:
    - o Each building starts as a disabled object.
    - o If the player selects a response, the algorithm modifies the object properties (e.g., texture, shape).
    - o It ensures a persistent world transformation, where choices have consequences for the outcome of the city blueprint and design.
- Pseudocode Presentation:

```
def update_scene(player_choice):
  for object in scene_objects:
    if is_affected_by_choice(object, player_choice):
      modify_object_properties(object, player_choice)
      render_updated_scene()
```

3.2.3. Player Decision Processing Algorithm

The main goal of the E-polis platform was to capture and process player decisions to support the fundamental research objectives of the game. Since each decision reflects potential sociopolitical or urban design preferences, a complex back-end system was required to orchestrate all related processes, including the collection, storage, and organization of information for later analysis. This algorithm was implemented to ensure the integrity of recorded responses while simultaneously preserving essential metadata, such as timestamps and player positions. This structured approach provided a viable solution for enabling the future use of gameplay data in research-oriented statistical tools and for integration into machine learning data pipelines.

The following breakdown outlines the rationale behind our algorithm, its architecture, and how it supports both real-time and persistent logging during daily gameplay sessions as well as future analytical needs.

- Core principle:
  - Collects and logs player responses in a structured format (but as unstructured data).
  - Ensures consistency and integrity in response collection (definitions of wrong execution and try-catch blocks for failures and errors during db communication or server authentication).

- Purpose of implementation:
  - Supports sociological and political research by mapping player choices to categories.
  - Enables researchers to analyze trends and decision patterns.

- How the system works:
  - Each player response was initially stored in a structured CSV file and then expanded this operation and stored in a remote DB repository in Brussels (Firebase).
  - Metadata such as time taken to answer, player position, and scene details are recorded.

- These data points can later be processed using statistical clustering and machine learning models.
- Pseudocode Presentation:

```
def log_player_decision(player_id, dilemma_id, response):
    timestamp = get_current_time()
    log_entry = 
        f"{player_id},{dilemma_id},{response},{timestamp}"
    write_to_RemoteDB("player_responses.Db", log_entry)
```

This method is significant for data collection, as it allows tracking of urban design preferences based on in-game decisions.

3.2.4. Distributed Player State Synchronization Algorithm

For our system to advance from single to multiplayer functionality and enable a shared urban development experience in the game world, it became necessary to synchronize game states across different users and, consequently, different electronic devices. The distributed synchronization algorithm enabled real-time data exchange between all participants, ensuring that the city transformations triggered by one player were instantly reflected to all other players by updating the variables of their environments. Firebase was chosen as the cloud infrastructure due to its scalability, simplicity, and responsiveness. This system guaranteed a consistent multiplayer experience, allowing players to collaboratively shape the city space while at the same time maintaining data accuracy and avoiding conflicts. As such, this section presents the logic, communication flow, and technical implementation of this synchronization mechanism within the multiplayer mode of our game.

- Core principle:
    - Synchronizes game state across multiple players in real-time.
    - Uses Firebase cloud storage to ensure consistency in game state.
- Purpose of implementation:
    - Allows multiple players to influence the same city without inconsistencies.
    - Prevents data loss by storing results remotely.
- How the system works:

- o Player actions and choices are broadcast to Firebase.
- o Other players receive live updates reflecting new game conditions (changes in city structure).
- o Ensures that all participants experience the same urban transformation process.
- Pseudocode Presentation:

```python
def sync_state_to_cloud(player_id, game_state):
    firebase.update(f"game_states/{player_id}", game_state)

def retrieve_state_from_cloud(player_id):
    return firebase.get(f"game_states/{player_id}")
```

Multiplayer synchronization is critical in E-Polis to ensure all players experience the same evolving city.

3.2.5. Endgame Consensus-Based Voting Algorithm

At the end of each game session, the players are given the opportunity to evaluate the outcome of their collective decisions through a final voting process. This means that a special feature exists in the E-polis game, as it allows the players to understand public sentiment in a simulated urban planning scenario. This feature, i.e., the voting algorithm, not only aggregates individual opinions but also conceals the outcome from the players to prevent bias in their reflections on their choices. As such, whether playing solo or in multiplayer mode, this mechanism provides a structured yet non-intrusive way to gather feedback on the final city design.

In the following section, we provide in detail the steps to collect, rank, and preserve the endgame evaluations.

- Core principle:
  - o Aggregates player votes on the final city structure.
  - o Uses a weighted ranking system to determine overall satisfaction.
- Purpose of implementation:
  - o Allows players to reflect on the collective decisions made during gameplay(single player=1 player or else multiplayer 2 to 6 players per room).

- - Provides researchers with insights into public preferences regarding urban planning.
- How the system works:
  - Each player submits a final vote (like, dislike, neutral).
  - Votes are aggregated and stored in Firebase.
  - The final consensus rating is not displayed in the endgame summary so as not to affect the players' decisions and perception of the final structure of the city.
- Pseudocode Presentation:

```python
def calculate_final_votes(votes):
  total_votes = len(votes)
  positive = sum(1 for v in votes if v == "like")
  negative = sum(1 for v in votes if v == "dislike")
  other    = sum(1 for v in votes if v == "other")
  consensus_score = [positive,negative,other]
  return consensus_score
```

The final voting mechanism allows players to rate the final city layout based on collective choices.

3.2.6. Game Workflow Algorithm

The final game workflow algorithm serves as the framework, i.e., the middleware layer that ties together the various subsystems of the E-polis game. This middleware workflow is responsible for maintaining continuity of operation between dilemmas, logging decisions, updating the city structure, and managing transitions through different phases of the gameplay. Importantly, this system ensures that no information is missed and that all player interactions are attributed correctly, whether this is the timestamp of an event or the event process itself. Moreover, it integrates cloud-based storage solutions and session tracking for robust multiplayer functionality. As such, this section explains how the workflow algorithm coordinates player input, system responses, and persistent data handling, and thus enables a seamless, research-ready gameplay experience.

- Core principle:
  - Ensures player interactions and decisions are stored in real-time.
  - Maintenance of accurate record of each player's responses and voting preferences.

- o Game state updates from building structures to scene transition to the final voting view from the above view.
- Purpose of implementation:
  - o To log and store players' responses to dilemmas.
  - o To track the game's progress and maintain session consistency.
  - o To preserve final voting results and the evolving city structure for later evaluation and analysis.
  - o To ensure data integrity and persistence remote database repository, preventing data loss.
- How the system works:
  - o The player submits an answer to an in-game dilemma.
  - o The system captures key metadata:
    - Player ID (Unique identifier).
    - Dilemma ID (Question being answered).
    - Selected Answer (Choice made by the player).
    - Timestamp (When the decision was recorded).
    - Game Room (The session the player is part of).
  - o The system constructs a database entry with this information.
  - o The data is stored in Firebase (a cloud-based game).
  - o If necessary, the system updates the game state based on the recorded response.

## 4. Results

In this section, we provide a detailed overview of the Python scripts and data-cleaning methods utilized in the development of our game. The process begins with the importation of the given .csv file, parsing it using the panda's library, and subsequently creating the two necessary tables for our database based on the categorized dilemmas explained in the previous section.

For a more comprehensive explanation and to facilitate future researchers, we have created a .ipynb file—a Python notebook (also known as a Jupyter notebook), [63]. This notebook contains the code, execution results, required libraries, and all other settings incorporated into our project to ensure correct execution. The file is available upon request to the corresponding author. It's important to note that we chose the .ipynb format over a .py file (plain text Python file) to enhance readability and ease of replication of results, as the notebook includes both code and execution outcomes, [64,65].

Furthermore, it is highlighted that in subsequent sections, we will introduce an analysis based on dummy data. Due to the software phase of the game and GDPR restrictions, we are unable to provide actual gameplay data without proper authorization, [66,67]. The data provided enhances the practicality of presenting the analysis while adhering to data protection regulations.

*4.1. Step 1: Import Python Library components*

In the development of our game and the associated analysis, we utilized several Python libraries for various tasks. Here's an overview of the key libraries incorporated into our project:

- <u>Pandas</u>: Used for data handling and manipulation of data frames (tabular data structures) for ETL (Extract, Transform, Load) operations on the provided .csv files containing questions and quests[3].
- <u>SQLite3</u>: Employed to define the database library (SQLite) for database interactions. This includes creating tables and performing CRUD (Create, Read, Update, Delete) operations[4].
- <u>NumPy</u>: Utilized for mathematical calculations and data analysis, providing functionality for defining and performing operations with arrays and matrices. It plays a crucial role in conducting mathematical operations on our game data[5].
- <u>Matplotlib</u>: Used to create plots and charts representing players' answers. Matplotlib is a versatile plotting library that supports a wide range of visualization types[6].
- <u>Plotly Express</u>: Similar to Matplotlib, Plotly Express extends our visualization capabilities, offering an extensive set of creative and interactive visualizations, especially for scatter plots. It provides additional features like annotations and legends on figures[7].
- Warnings: We employed "warnings.simplefilter(action='ignore', category=FutureWarning)" to suppress warnings during the execution of the Jupyter Notebook. This ensures a smoother execution flow and helps in handling errors and messages more efficiently[8].
- Time: Used to generate timestamps from the local work machine's execution time. Timestamps are incorporated into the generated output files of our tests, providing a temporal reference for analysis[9].

These libraries collectively contribute to the efficiency and functionality of our game development and subsequent data analysis, enabling a seamless workflow in handling, processing, and visualizing the game data.

---

[3] Pandas. Pandas *Documentation* **2025**. Available online: https://pandas.pydata.org/ (accessed on 19 May 2025).

[4] Python Software Foundation. SQLite3 Documentation. Python Software Foundation **2025**. Available online: https://docs.python.org/3/library/sqlite3.html (accessed on 19 May 2025).

[5] *NumPy*. **2025**. Available online: https://numpy.org/ (accessed on 19 May 2025).

[6] *Matplotlib*. **2025**. Available online: https://matplotlib.org/ (accessed on 19 May 2025).

[7] *Plotly Express*. **2025**. Available online: https://plotly.com/python/plotly-express/ (accessed on 19 May 2025).

[8] *Python Software Foundation.* Python Warnings. **2025**. Available online: https://docs.python.org/3/library/warnings.html (accessed on 19 May 2025).

[9] *Python Software Foundation*. Python Time. Time command **2025**. Available online: https://docs.python.org/3/library/time.html (accessed on 19 May 2025).

*4.2. Step 2: Parse the Csv, Analyze Data, Create Db Tables*

In handling a given CSV file, such as "data.xls," following discussions with the team delivering the dilemmas, we meticulously review the document's specific columns. Precisely, guided by the provided input, we establish a structured arrangement of rows and columns for every identified dilemma group, as outlined in our case study input. Subsequently, we proceed to formulate the concrete foundation of our database.

Each set of questions, derived from distinct Excel column groups, prompts the creation of a dedicated table. To achieve this, three tailored functions have been devised to execute specific processes. These functions facilitate the seamless scanning, extraction, and construction of tables in alignment with the unique requirements of each dilemma group. This methodical approach ensures the systematic organization and incorporation of the received dilemmas into our database, contributing to the overall coherence and effectiveness of the data management processes:

- create_table function accepts a data frame, selects specific rows and columns, and uses an existing DataFrame to perform the selection. Subsequently, it employs the iloc method to extract the chosen rows and columns from the original DataFrame, resetting the index of the new DataFrame to start from zero.
- create_table_diag function rearranges data into a diagonal pattern to generate a new DataFrame. While similar to create_table, this function creates a distinct data structure. It assembles a new DataFrame from an existing one by organizing the data diagonally. To achieve this, the function calculates the dimensions of the original DataFrame (N, M), forms a list of column names for the new DataFrame (including original column names and diagonal column names), and populates the table with zero values except for the diagonal entries.
- create_db function takes two DataFrames and stores them in an SQLite database with two tables. Initially, it establishes a connection to the SQLite database. Subsequently, it utilizes the to_sql method to save the two DataFrames to the database.

4.2.1. Example of the actual dataset with player's dilemmas

In Figure 3 and Figure 4, we present an actual representation of the tables created based on this step for each scenario within the dilemmas group (the answers are in Greek):

**Figure 3.** Actual representation of the table of the Db for the first group of dilemmas (Political Philosophy Dilemmas).

**Figure 4.** Actual representation of the table of the Db for the second group of dilemmas (International Relations and Political Theory Dilemmas).

4.2.2. Example of the actual dataset with player's dilemmas in a diagonal pattern

Similarly, in Figure 5 and Figure 6, if we switch the structure of our database to have the answers to specific venues stored along the diagonal of our table, the resulting structure would be as follows (the answers are in Greek):

**Figure 5.** Actual representation of the diagonal table of the Db for the first group of dilemmas (Political Philosophy Dilemmas).

**Figure 6.** Actual representation of the diagonal table of the Db for the second group of dilemmas (International Relations and Political Theory Dilemmas).

This data structure is not currently in use, but we have created it for future reference by other teams in the research project. Specifically, data scientists may utilize it as a one-hot encoding in case the questions change or expand, implying an ordinal relationship. In this setup, instead of having 0 or 1, we have placed the answer for each dilemma group along the diagonal. Additionally, this data can be linked with the player's position or the level (scene) structures (buildings, etc.). This linkage allows it to be associated with specific coordinates that alter the structure of the buildings and game objects, thereby simulating the user experience and dynamically changing the player's position and graphics throughout the gameplay.

*4.3. Step 3: Analyzing Data and Export DB Data*

Table 1 within our database encompasses a diverse set of questions, each accompanied by six answer options that collectively represent a broad spectrum of potential responses. In contrast, Table 2 focuses on a more specific set of questions, each paired with the same six answer options of different profiling, providing a targeted assessment tool. During gameplay, player responses are stored in our database.

As the data is categorical, statistical measures like averages or standard deviations are inapplicable. Our focus shifts to visualizing and comparing the data to gain insights into response distribution and patterns within the answer options. This approach offers a comprehensive understanding of the data's characteristics, guiding future research.

To facilitate this, we've created the save_answer function, tailored to store survey data in a SQLite database. It takes four parameters: array (data to be saved), questions (column names), table (defaults to 1, specifying the table number), and n (an optional additional identifier).

The function converts the array into a pandas DataFrame using specified column names. It generates a timestamp string, creating a unique filename for the SQLite database. The connection to the SQLite database is established using this filename. The resulting output is a SQLite database file with a filename format like "Db1Table{table}{n}{timestr}.sqlite" or "Db1Table{table}_{timestr}.sqlite" if n is not provided. This database contains a table named "answer" with the survey data. The same methodology is applied to the second data structure with dilemmas arranged in a diagonal pattern.

*4.4. Step 4: Visualizing Responder's Data*

In this section, upon completion of all available quests by the player, we generate insightful graphs to comprehend their profile and behaviour. As presented in Figure 7, the phone booth in each level enables the player to advance to the next level or conclude the game, and to reach it he/she must have answered all available questions.

**Figure 7.** The exit from each level of our game is performed via going in the phonebooth where for the player to reach it he/she must have answered all available quests.

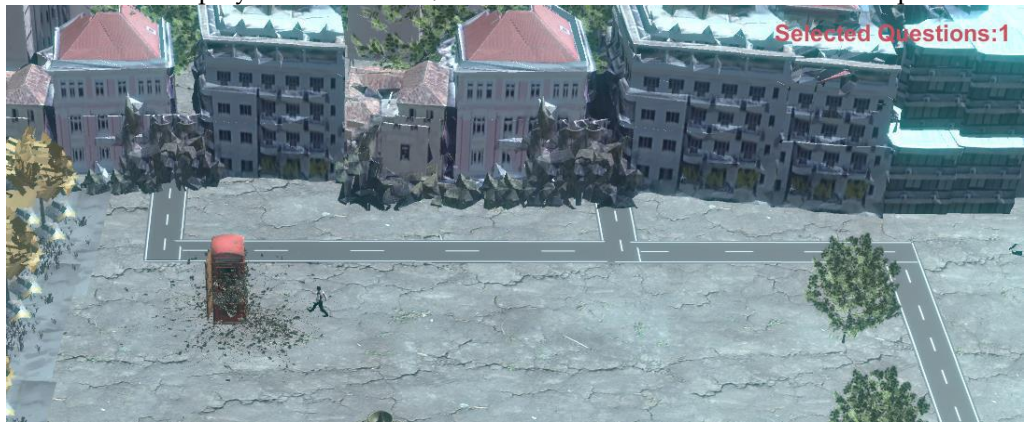

The following figures consist of histograms, (Figure 8 and Figure 9) corresponding to each group and polar "spider" plots, commonly known as radar plots (Figure 10 and Figure 11) where the answers are in Greek. Specifically, if we study Figure 9 and the histogram plots of these data, the distribution suggests that certain ideologies, e.g. Realism and Humanism, are more intuitively selected, possibly because their descriptions resonate more clearly or appear less extreme to players, thus helping us track and monitor a design bias.

Similarly, in regards to Figure 11, the symmetry or asymmetry of the spider plots allows us to detect ideological polarization or openness in a player's profile, with spiked axes indicating strong leaning and flat ones suggesting neutrality or indecision. Similarly, Figure 12 provides a reference for validating the effectiveness of the questions posed, i.e. a uniform spread suggests neutrality whereas a skewed distribution may imply a persuasive or biased pattern.

**Figure 8.** Indicative Histogram plot for the first group of dilemmas (Political Philosophy Dilemmas).

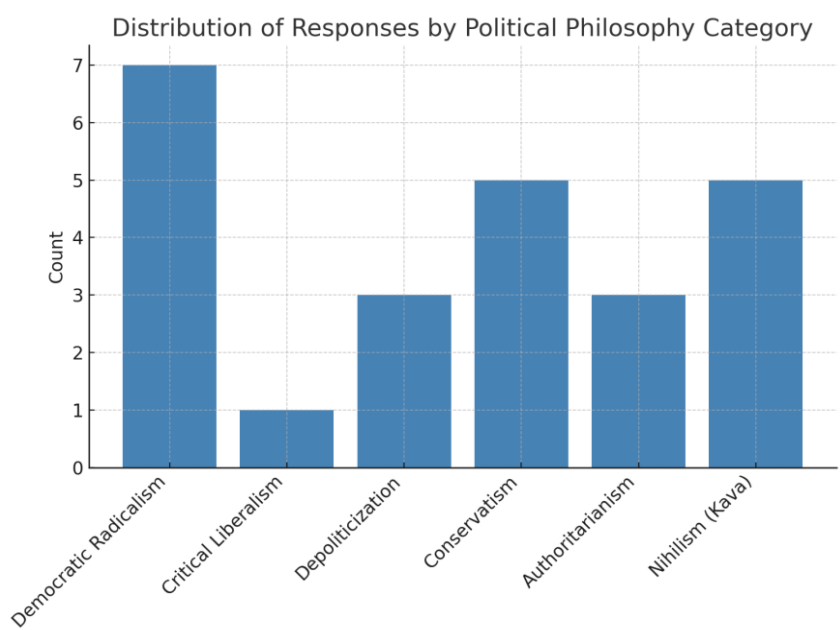

The exit from each level of our game is performed via going in the phonebooth where for the player to reach it he/she must have answered all available quests.

**Figure 9.** Indicative Histogram plot for the second group of dilemmas (International Relations and Political Theory Dilemmas).

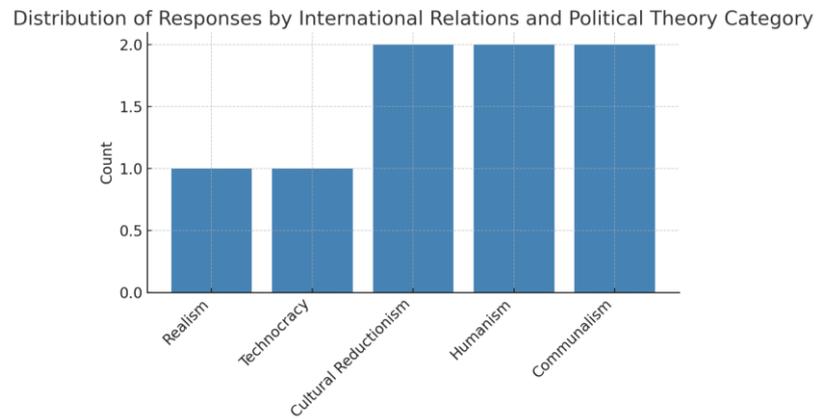

**Figure 10.** Indicative Polar "spider" plot for the first group of dilemmas (Political Philosophy Dilemmas).

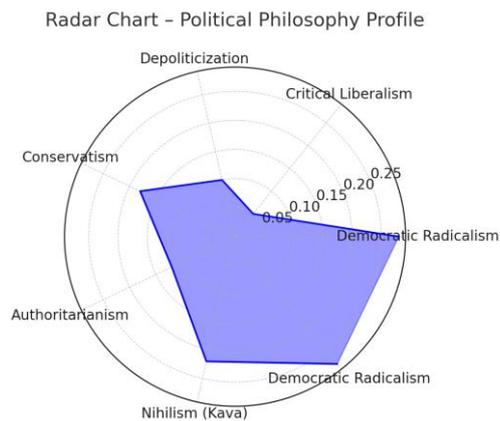

**Figure 11.** Indicative Polar "spider" plot for the second group of dilemmas (International Relations and Political Theory Dilemmas).

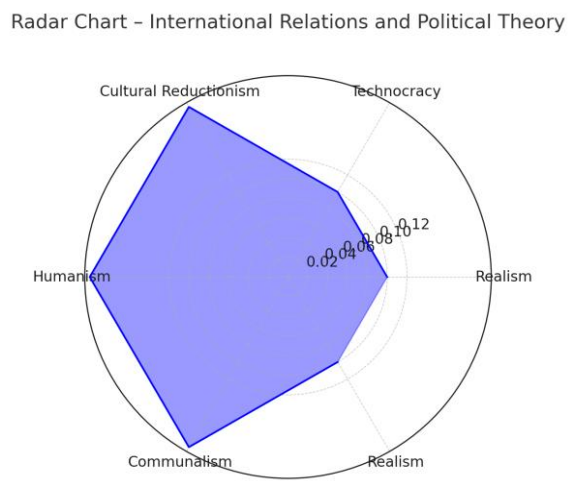

The stack bar illustrations of our study are presented in Figure 12 and Figure 13 below (the answers are in Greek). In these Figures, the x-axis represents the different questions from your survey, labelled as "Q1", "Q2", and so on. The y-axis represents the count of responses for each answer option. The text annotations on each cell represent the count of responses for each answer option of each question (response distribution). As such, avoidance of specific ideologies, such as Cultural Reductionism, may reflect either a conscious distancing from controversial views or a lack of clarity in how these options are presented.

**Figure 12.** Indicative Stack bars for the first group of dilemmas (Political Philosophy Dilemmas). Specifically, the blue colour represents Democratic Radicalism (ΔΗΜΟΚΡΑΤΙΚΟΣ ΡΙΖΟΣΠΑΣΤΙΣΜΟΣ), the orange represents Critical Liberalism (ΚΡΙΤΙΚΟΣ ΦΙΛΕΛΕΥΘΕΡΙΣΜΟΣ), the green represents Depoliticization (ΑΠΟΠΟΛΙΤΙΚΟΤΗΤΑ), the red represents Conservatism (ΣΥΝΤΗΡΗΤΙΣΜΟΣ), the purple represents Authoritarianism (ΑΥΤΑΡΧΙΣΜΟΣ), and the brown represents Other/Unclassified (ΚΑΒΑ).

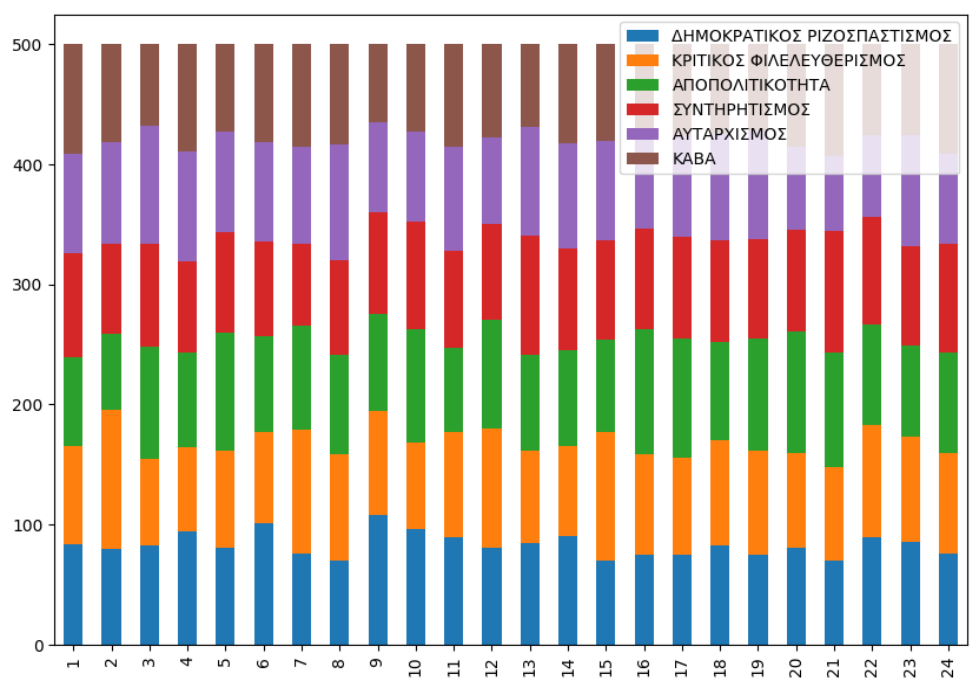

**Figure 13.** Indicative Stack bars for the second group of dilemmas (International Relations and Political Theory Dilemmas). Specifically, the blue colour represents Realism (ΡΕΑΛΙΣΜΟΣ), the orange represents Technocracy (ΤΕΧΝΟΚΡΑΤΙΑ), the green represents Cultural Reductionism (ΠΟΛΙΤΙΣΜΙΚΟΣ ΑΝΑΓΩΓΙΣΜΟΣ), the red represents Humanism (ΑΝΘΡΩΠΙΣΜΟΣ), the purple represents Meritocracy (ΑΞΙΟΚΡΑΤΙΑ), and the brown represents Communitarianism (ΚΟΙΝΟΤΙΣΜΟΣ).

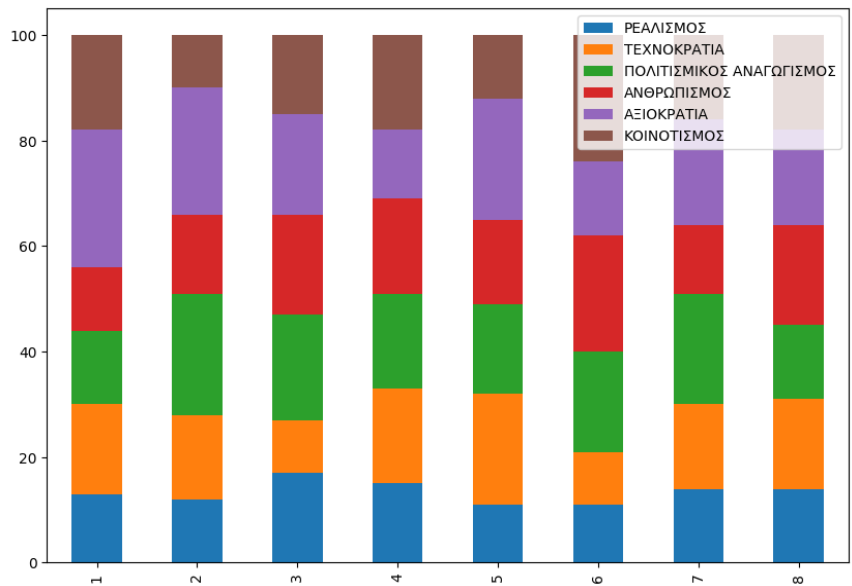

Similarly, the heatmap illustrations based on the same datasets of the previous images are presented in Figure 14 and Figure 15 below (the answers are in Greek). In these figures, the x-axis represents the questions ("Q1", "Q2", etc.), and the y-axis represents the response categories. The text annotations on each cell represent the count of responses for each category of each question. As such, the sparsity in some cells of the heatmap is important as it may point to cognitive fatigue or reduced motivation in later-stage questions, suggesting a potential need for enhancing the overall gamification experience (adaptive pacing or periodic gameplay breaks).

**Figure 14.** Indicative heatmap graph for the first group of dilemmas (Political Philosophy Dilemmas). Specifically, the first line represents Democratic Radicalism (ΔΗΜΟΚΡΑΤΙΚΟΣ ΡΙΖΟΣΠΑΣΤΙΣΜΟΣ), the second line represents Critical Liberalism (ΚΡΙΤΙΚΟΣ ΦΙΛΕΛΕΥΘΕΡΙΣΜΟΣ), the third line represents Depoliticization (ΑΠΟΠΟΛΙΤΙΚΟΤΗΤΑ), the fourth line represents Conservatism (ΣΥΝΤΗΡΗΤΙΣΜΟΣ), the fifth line represents Authoritarianism (ΑΥΤΑΡΧΙΣΜΟΣ), and the sixth line represents Other/Unclassified (ΚΑΒΑ).

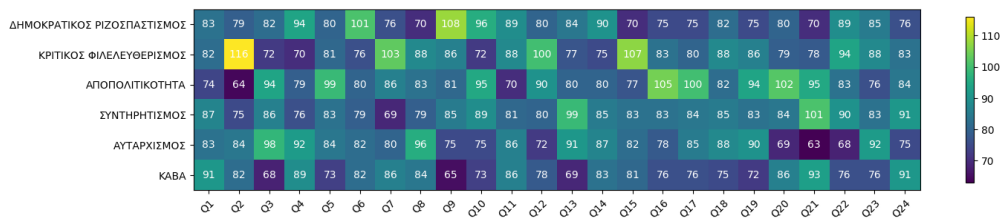

**Figure 15.** Indicative heatmap graph for the second group of dilemmas (International Relations and Political Theory Dilemmas). Specifically, the first line represents Realism (ΡΕΑΛΙΣΜΟΣ), the second line represents Technocracy (ΤΕΧΝΟΚΡΑΤΙΑ), the third line represents Cultural Reductionism (ΠΟΛΙΤΙΚΟΣ ΑΝΑΓΩΓΙΣΜΟΣ), the fourth line represents Humanism (ΑΝΘΡΩΠΙΣΜΟΣ), the fifth line represents Meritocracy (ΑΞΙΟΚΡΑΤΙΑ), and the sixth line represents Communitarianism (ΚΟΙΝΟΤΙΣΜΟΣ).

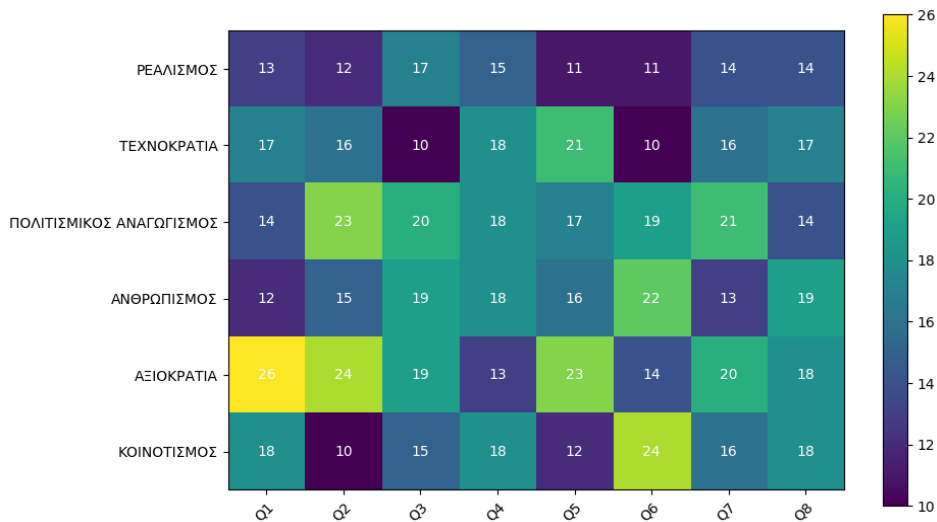

*4.5. Step 5: Validate the ground truth of our results: PCA analysis*

Moreover, based on the above, for a dataset, we have generated data responses to perform a detailed PCA analysis and then attempted to cluster the data using K-means and other algorithms. Based on our data analysis, a common method to determine the optimal number of components is the "Elbow Method," which involves plotting the explained variance ratio and selecting the number of components at the elbow point. However, in our case, we set the number of components to 3 for visualization purposes, although our code calculates all the necessary priority values.

For example, to fit PCA on scaled data, we plotted the cumulative summation of the explained variance in Figure 16. Then, we applied our PCA, i.e., transformed the data to reduce the number of components to 3, to print our results to the console, visualize them, and understand them more easily. Specifically, we used K-means clustering and agglomerative clustering algorithms, as suggested by the scikit-learn Python package documentation, and extracted the results shown for K-means clustering in Figure 16, Figure 17, Figure 18, Figure 19 and Figure 20. It is noted that the clustering patterns in PCA 2D and 3D views support the hypothesis that players fall into latent behavioural groups, offering a quantitative basis for later classification models or personalization strategies.

**Figure 16.** PCA Cumulative Summation of the Explained Variance

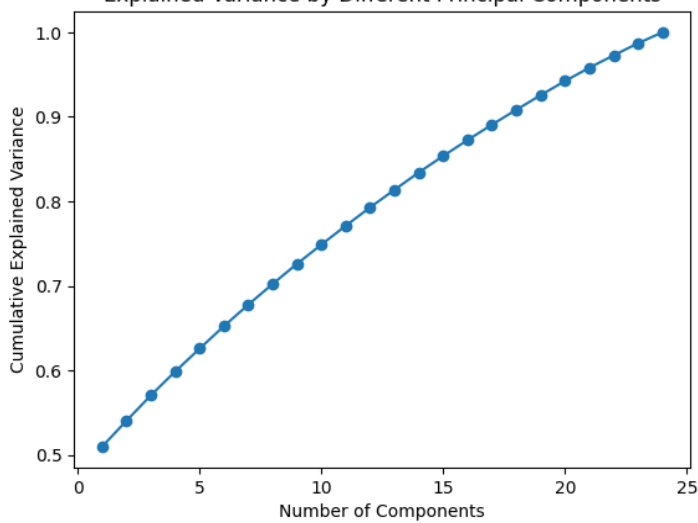

**Figure 17.** Clustering representation using Kmeans for the dataset samples in 3d

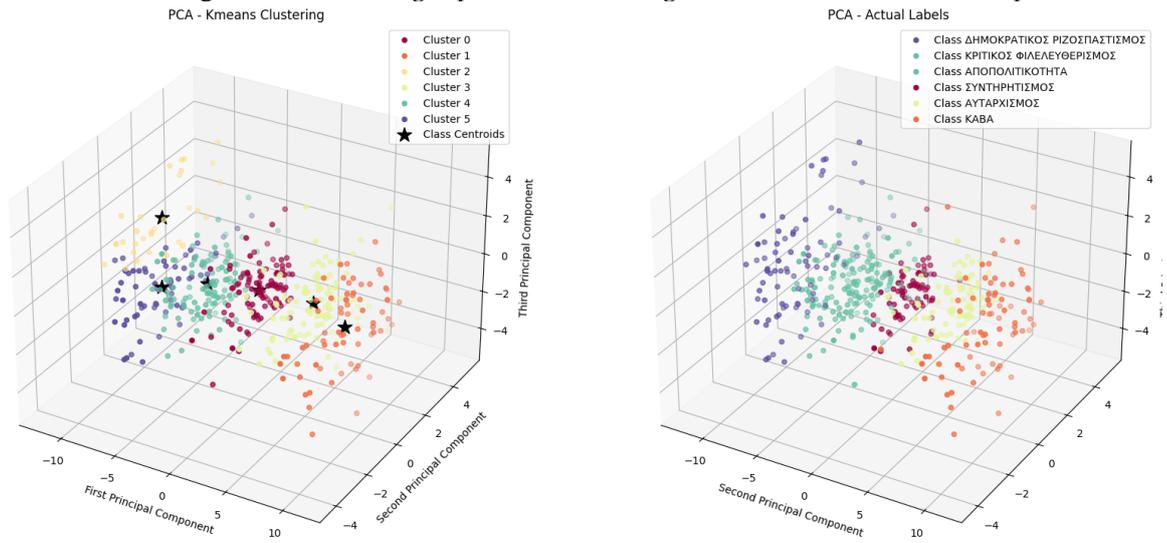

**Figure 18.** Clustering representation using Kmeans for the dataset samples in 2d

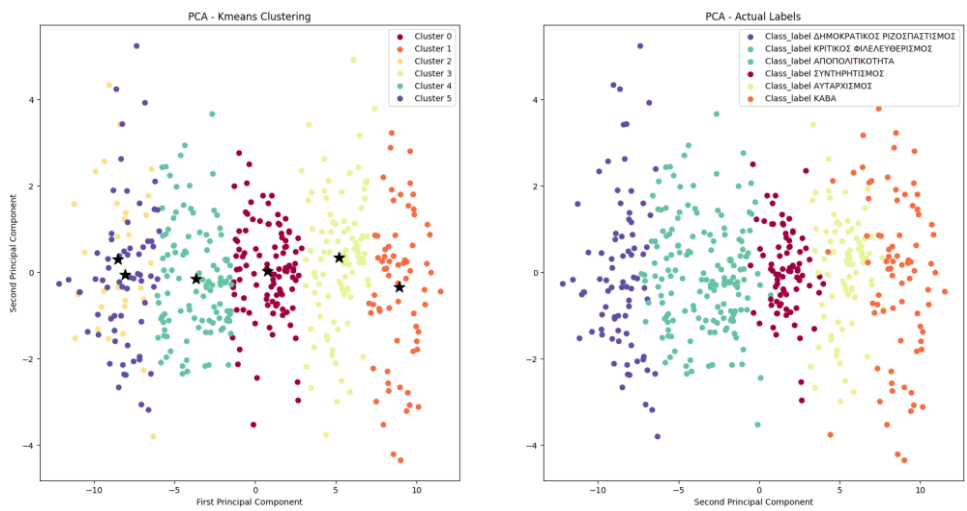

**Figure 19.** Clustering representation using Agglomerative for the dataset samples in 3d

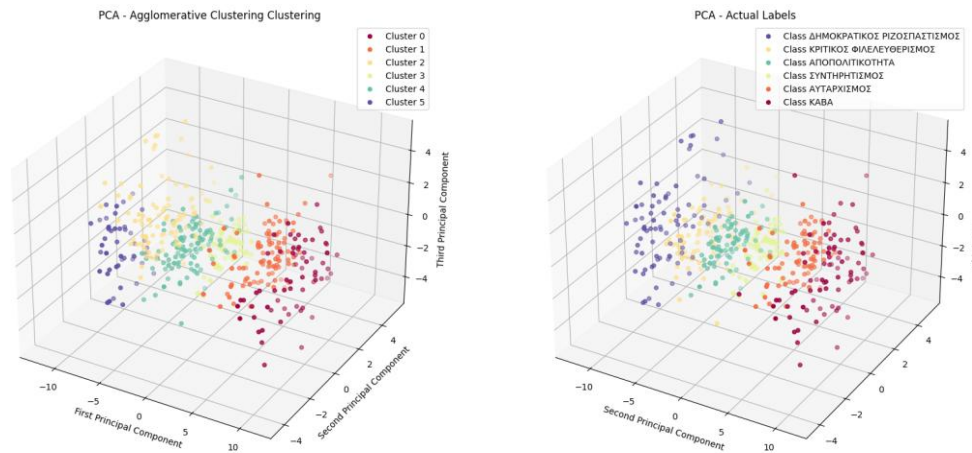

**Figure 20.** Clustering representation using Agglomerative for the dataset samples in 2d

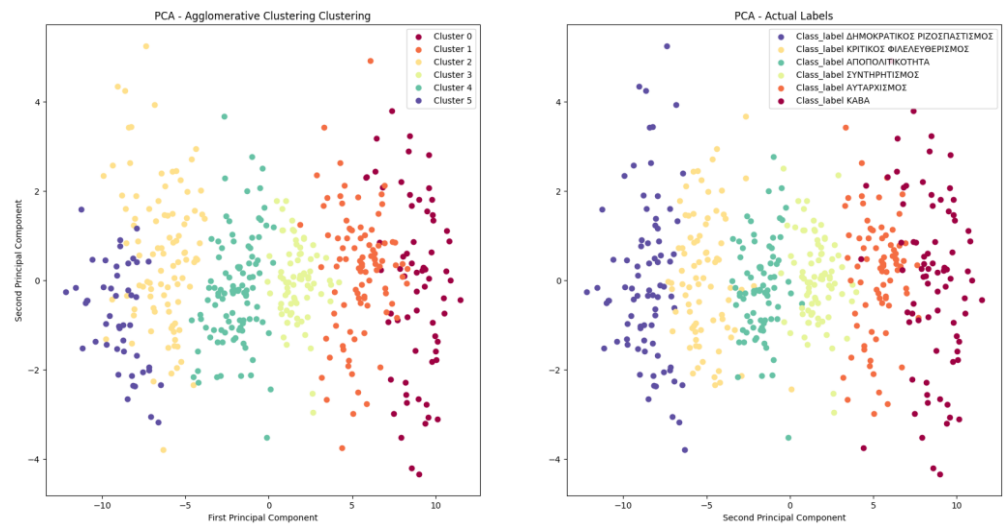

## 5. Discussion

### 5.1. Data Analysis

The analysis of the dataset used showcased recurring patterns, deviations, and simulated behavioural traits reflecting how players might respond to sociopolitical dilemmas in real simulated scenarios. Beginning with the histogram plots for the Political Philosophy Dilemmas (Figure 8), a clear dominance of Democratic Radicalism (blue) and Critical Liberalism (brown) emerged across most questions, suggesting that the dummy respondents were designed to reflect a progressive stance, often aligned with youth demographics in real-world surveys. Similarly, in Figure 9, visualizing responses to the International Relations and Political Theory Dilemmas, preferences were concentrated on Realism (blue) and Humanism (red), reflecting pragmatic yet ethically inclined profiles. These preferences were reinforced in the corresponding polar (spider) plots (Figures 10 and 11), suggesting a stronger alignment or repeated selection.

Beyond dominant patterns, several anomalies were also observed. In the stacked bar charts (Figures 12 and 13), these ideologies suggest a societal avoidance of these extreme or marginalized positions. The heatmaps (Figures 14 and 15) reveal sparse or missing data points, particularly around questions Q7 to Q10, suggesting either reduced engagement with late-stage dilemmas or fatigue in longer-game- sessions. If replicated with another dataset of more players or taking into account a study of players from different age or cultural backgrounds, these trends could imply difficulty in answering complex or abstract questions toward the end.

In terms of player behaviour traits, Figures 10 and 11 demonstrate two distinct profiles: high polarization, favouring one ideology across all dilemmas, forming skewed radar plots, and balanced or exploratory patterns, creating circular or flower-like spider plots. These signatures could classify users into categories like "ideologically consistent," "deliberative," or "experimental." The stacked bar plots (Figures 12–13) show clustering of preferences by question, implying that certain questions consistently evoke specific answers, likely due to question framing or scenario context.

In addition, it is worth mentioning that each figure not only visualizes the preferences and tendencies of players but also reveals insights into the decision-making process. For instance, the histogram plots (Figure 8 and Figure 9) can be used as a means to suggest an ideological distribution, or even a comparison overview framework thus highlighting potential biases toward ideologies. Similarly, the polar plots (Figure 10 and Figure 11) reinforce this by spatially clustering ideologies, offering a quick visual representation of ideological coherence. Meanwhile, the stacked bar charts (Figure 12 and Figure 13) reveal how individual questions align with particular ideologies, thus implying either strong contextual framing or inherent ideological triggers in the questions themselves. Lastly, it is noted that the heatmaps (Figure 14 and Figure 15) are particularly useful for spotting underrepresented choices and engagement gaps across the timeline of gameplay, which can inform revisions in question ordering or user interface to reduce fatigue and abandonment.

Collectively, these visualizations and their embedded patterns highlight the analytical potential of the E-polis framework. Given a better or better dataset of participants, the layered representations—histograms, radar plots, stacked bars, and heatmaps can offer a robust foundation for interpreting sociopolitical behaviour. The anomalies, when intentionally generated, provide a sandbox for testing edge cases and refining data collection and analytical accuracy in future gameplay with real participants. The suggested statistical analysis visualization can either be used to explain a session of gameplays by players of specific characteristics or used as a means to validate the ground truth of classification by other games/activities/questionnaires or even AI tools used to determine their -the players- properties.

These visualization tools act as real-time feedback mechanisms that enhance the final design of serious games. For developers and social scientists alike, observing and monitoring players' response distributions through heatmaps or stacked bar charts is important, as it may indicate patterns of cognitive overload or help adjust difficulty levels and narrative pacing to improve the gamification experience. Additionally, comparative analysis across different demographic groups can reveal how sociopolitical backgrounds influence navigation through ideologically charged dilemmas. In this way, the game's tools serve as a mirror of broader social tendencies—revealing not only player preferences but also feelings of discomfort, confusion, or disengagement. Therefore, extending these visualizations with additional layers—such as time spent per question—could enhance the interpretability and utility of the data layer within the middleware structure. In future iterations, these visual features could assist in the real-time adaptation of our game. This means that it can be used as a tool to guide the placement or sequencing of dilemmas to maintain player engagement and maximize data quality.

*5.2. Limitations*

5.2.1. Game Middleware – Large-scale architectural design

The game middleware solutions also have critical limitations, mainly the lack of scalable data management, the inability to adapt to user behaviour dynamically and the constraints of the traditional client-server model in game development operations. Many game middleware frameworks are not designed for large-scale, multiplayer environments, where significant amounts of player-generated data must not only be collected, analyzed, and stored but also used to shape the game world and player choices. Traditional game engines primarily focus on rendering and physics simulations, offering neither a structured data-handling system nor built-in mechanisms for integrating external databases (e.g., Unity, one of the most widely used game engines, supports C# but lacks out-of-the-box integration for common databases like Microsoft SQL Server or MySQL).

Our game middleware approach addresses this limitation by applying IoT middleware principles to game development. Specifically, our game middleware treats player interactions as real-time data streams, similar to sensor networks. This allows for dynamic scene management and efficient storage of player data, making the game both scalable and data-driven.

### 5.2.2. Game Middleware – Adaptive architectural design principles

Regarding adaptation to user data points, most game middleware solutions rarely incorporate adaptive mechanisms based on player behaviour. They merely collect user interactions but fail to utilize them dynamically, for instance, for real-time scene modifications or personalized content delivery.

To address this, our middleware integrates a smart scene transition mechanism, which adapts game content dynamically based on player-generated data. This means that the proposed game development middleware treats the game as a data-producing smart sensor, enabling AI-driven procedural content generation that enhances player immersion and engagement.

### 5.2.3. Game Middleware – Traditional synchronous client-server communication

Nowadays, the features of a traditional client-server, request-response model are relatively limited. This is the case because clients may only send data after they submit their request to the server. This situation limits developers in terms of creating dynamic applications.

Our game middleware approach addresses this issue by implementing event-driven communication rather than relying on synchronous requests. This means that instead of requiring clients to explicitly send requests before receiving data, our middleware treats each player and their interactions as part of a sensor network with a unified real-time data stream. This allows for continuous data exchange and instantaneous state synchronization, making it more suitable for dynamic multiplayer environments.

Lastly, this feature is further supported by our smart scene transition mechanism, which dynamically adapts game content based on real-time player behaviour without requiring constant back-and-forth communication with a centralized server. By leveraging decentralized processing techniques, our system optimizes latency, ensuring that gameplay remains seamless and responsive, even under heavy computational loads.

### 5.2.4. IoT & Game Middleware – Security & Service Management in IoT SOA Middleware

One of the key issues in a game middleware layer that supports multiplayer is ensuring that communication, access control, and service management are effectively handled within a Service-Oriented Architecture (SOA). Traditional middleware solutions often focus on persistent client-server connections but lack a structured approach for authorizing access to different middleware layers, modules, services, and components. Moreover, middleware must act as a secure data transmission layer, managing encrypted communication (e.g., SSL Handshakes) and diagnosing system failures while maintaining low latency and high availability. Without these features, middleware fails to fully integrate security, reliability, and real-time communication, making it vulnerable to unauthorized access, inefficient service distribution, and limited scalability in multi-device environments.

Our middleware addresses these issues three-fold:

- Extends a SOA-based communication layer by treating players and their interactions as (real-time) data streams, similar to sensor networks.
- Instead of traditional client-server connections, ensures event-driven synchronization between game instances, dynamically managing service requests and state transitions without requiring continuous polling.
- Introduces a smart scene transition mechanism that ensures data is securely transmitted and validated before affecting game state changes, thus preventing unauthorized client manipulations.

## 6. Conclusions

Our game development process and various layers have been extensively discussed in other works. Therefore, we have focused our efforts on elucidating the Game layer, which is responsible for the actual game and from which we gather valuable data points. The study primarily concentrates on outlining a preliminary analysis and the creation of a database from players' gameplay. It provides a detailed top-down approach to extracting data from a CSV, structuring it for loading into a database, and subsequently visualizing the data. Additionally, we propose a diagonal-shaped tabular format that can serve as a one-hot encoding structure in a neural network. This format can also be utilized to establish a grid of points between the values of the diagonal and the actual game object coordinates. This enables dynamic changes to the game graphics during runtime, adjusting player and graphic elements' positions on a set grid or terrain.

As such, the data visualization and dimensionality reduction presented in Figures 8 through 20 offer some preliminary but meaningful insight into player behaviour throughout the game and, most importantly, their political choices, thus revealing their orientation within a serious game environment. The histogram plots (Figures 8 and 9) depict the frequency distribution of the players' responses for each category, revealing biases toward particular ideologies (e.g., higher prevalence of selection under democratic radicalism or technocracy) and also showcasing the diversity and uniformity of player perspectives from different sociopolitical contexts.

As such, the polar spider plots (Figures 10 and 11) that were presented assist in our understanding by spatially distributing the same categorical data around a central axis, allowing for an intuitive comparison of our six study political profiles. The resulting plot geometry—whether balanced, skewed, or spiked—offers a visual fingerprint of the political mindset we situate in this manuscript. Similarly, stacked bar charts and heatmaps (Figures 12 to 15) showcase aggregated responses, not as individual item-level response distributions but to suggest actual population-level tendencies. For example, a dominant presence of certain colours across different questions indicates a recurring ideological leaning or cognitive bias embedded in the population or a preset dataset.

Similarly, the heatmaps provide a compact summary of frequency intensities, highlighting which responses track ideological clustering and determining the sociopolitical richness of the dataset studied. As a result, we conclude our research using PCA and clustering results (Figures 16 to 20) to provide a quantitative dimension to these qualitative visualizations. Analytically, by reducing the high-dimensional response space to three components, PCA enables efficient comparison of variance across game users. The scatter plots and cumulative explained variance help illustrate the formation of clear user clusters using both K-means and agglomerative methods. This means that these clusters may suggest distinguishable sociopolitical profiles among players, and their underlying ideological variables can be used for future and potential classification, interaction, and voting outcomes.

Looking ahead, the authors aspire that this analysis of the steps and the elucidation of how this layer mines and visualizes data based on a CSV file will aid future researchers, especially those specializing in social surveys. For future endeavours, it is imperative to test and evaluate whether the envisioned addition of Virtual Reality (VR) elements to the game influences players' actions. It would be intriguing to observe players engaging with and without VR goggles to discern potential behavioural changes. Similarly, one can use this data as a means to provide AI to implement the potential for typological classification, decision patterns, and, based on scene interaction and in-game behaviour, voting outcomes. In this regard, the data pipeline—from response recording to statistical clustering—demonstrates, to the best of our knowledge, a strong potential for empirical psychological exploration through serious game frameworks. It is crucial to acknowledge that this analysis is not definitive but rather serves as a method to explore certain quality attributes or validate aspects of deep data analysis, acting as a means to validate the ground truth method.

## 7. Appendix

*Game Workflow Pseudocode Presentation*

The final voting mechanism allows players to rate the final city layout based on collective choices as presented in the pseudocode below:

```python
import time
import firebase_admin
from firebase_admin import credentials, db

# Firebase Authentication Setup
 def initialize_firebase():
 """Initializes Firebase connection using service account credentials. Ensures secure authentication and access to the database. """
  cred = credentials.Certificate("path/to/serviceAccountKey.json")
  firebase_admin.initialize_app(cred, {
                        'databaseURL':
                    'https://our-database-name.firebaseio.com/'
                         })
 print("Firebase initialized successfully.")

 # Function to get current timestamp
 def get_current_timestamp():
 """Returns the current timestamp in a standard format."""
  return time.strftime('%Y-%m-%d %H:%M:%S')

 # Store player response in Firebase
 def store_player_answer(player_id, dilemma_id, selected_answer, game_room, auth_token):
 """Stores the player's response securely in Firebase
(param player_id: Unique identifier for the player,
param dilemma_id: Unique identifier for the dilemma
question, param selected_answer: The player's chosen
```

```python
response, param game_room: The game session the player
is part of, param auth_token: Authentication token for
verifying user access."""
    timestamp = get_current_timestamp()
    # Authenticate user session
    if not authenticate_player(player_id, auth_token):
        print(f"Authentication failed for Player {player_id}. Data not stored.")
        return

    # Reference to Firebase game room answers
    ref = db.reference(f'game_rooms/{game_room}/player_answers')
    ref.push({
        'player_id': player_id,
        'dilemma_id': dilemma_id,
        'answer': selected_answer,
        'timestamp': timestamp
    })

    print(f"Stored response for Player {player_id} in Room {game_room}")

# Firebase Authentication Function
def authenticate_player(player_id, auth_token):
    """Validates the player's authentication token before storing data (where, param player_id: Unique identifier for the player, param auth_token: Token issued for verifying player authentication, return: Boolean indicating whether authentication was successful."""
    auth_ref = db.reference(f'authenticated_users/{player_id}')
    stored_token = auth_ref.get()

    if stored_token == auth_token:
        return True
    else:
        return False
```

```python
# Initialize Firebase Connection
initialize_firebase()

# Example Usage
store_player_answer(player_id=12,    dilemma_id=5, selected_answer="Option  A",   game_room=2,   auth_token="valid_token_123")
store_player_answer(player_id=15,    dilemma_id=7, selected_answer="Option  C",   game_room=3,   auth_token="invalid_token_456")
```

**Supplementary Materials:** Not applicable.

**Author Contributions:** Both authors made significant contributions to this work. A.G. and E.K. conceived and designed the experiments. A.G. coded the application; was responsible for the investigation, methodology, software, validation, visualization, writing the original draft, and reviewing and editing resources; and carried out the simulation and the optimization and writing of—original draft. E.K. contributed to the conceptualization, investigation, methodology, project administration, resources, software, supervision, validation, visualization, review, and editing. All authors have read and agreed to the published version of the manuscript.

**Funding:** This research was funded by the Hellenic Foundation of Research and Innovation (H.F.R.I.), in the context of the "1st Call for H.F.R.I. (http://www.elidek.gr) Research Projects to Support Faculty Members & Researchers and Procure HighValue Research Equipment" (Project Number: 2617)

**Data Availability Statement:** The data presented in this study are available on request from the corresponding author.

**Acknowledgements:** The authors would like to thank Mr. Gerasimos Kouzelis for providing the research outline of this project and Mr. Orestis Didi for his overall assistance and expertise on the topic.

**Conflicts of Interest:** The authors declare no conflicts of interest